\documentclass[10pt,leqno]{amsart}
\usepackage{graphicx}
\pdfoutput=1
\usepackage{indentfirst,csquotes}

\usepackage{amsmath,amssymb,amsfonts}
\usepackage{algorithmic}
\usepackage{graphicx}
\usepackage{textcomp}
\usepackage{xcolor}
\def\BibTeX{{\rm B\kern-.05em{\sc i\kern-.025em b}\kern-.08em
    T\kern-.1667em\lower.7ex\hbox{E}\kern-.125emX}}

\usepackage{algorithm}
    
\begin{document}

\title[Extracting Patterns of Chemical Information from DMS]{Extracting Patterns of Chemical Information from Differential Mobility Spectrometry Measurements under Varying Conditions of Humidity and Temperature}
\author[Müller et al.]{Philipp~Müller$^{1*}$, Gary~A.~Eiceman$^2$, Anton~Rauhameri$^1$, Anton~Kontunen$^{1,3}$, Antti~Roine$^{1,3}$, Niku~Oksala$^{1,3,4}$, Antti~Vehkaoja$^{1}$, and~Maiju~Lepomäki$^{1,5}$}

\let\thefootnote\relax
\footnotetext{$^{*}$corresponding author: philipp.muller@tuni.fi\\
$^1$Sensor Technology and Biomeasurements, Faculty of Medicine and Health Technology, Tampere University, FI-33720 Tampere, Finland\\
$^2$Department of Chemistry and Biochemistry, New Mexico State University, Las Cruces, NM 88003-8001, USA\\
$^3$Olfactomics Ltd, Kampusareena, Korkeakoulunkatu 7, FI-33720 Tampere, Finland\\
$^4$Vascular Centre, Tampere University Hospital, Central Hospital, P.O. Box 2000, FI-33521 Tampere, Finland\\
$^5$Department of Pathology, Fimlab Laboratories, Arvo Ylpön katu 4, FI-33520 Tampere, Finland

Version: 23 February 2026

This work has been submitted to Springer for possible publication. Copyright may be transferred without notice, after which this version may no longer be accessible.
}

\maketitle

\begin{abstract}
Differential Mobility Spectrometry (DMS), also known as Field Asymmetric Ion Mobility Spectrometry, is a rapid and affordable technology for extracting information from gas phase samples containing complex volatile organic compounds, and can therefore be used for analyzing surgical smoke. One obstacle to its widespread application is the dependence of DMS measurements on humidity and, to a lesser degree, temperature, making comparison of data measured under different environmental conditions arbitrary. The commonly used solution is to regulate these environmental conditions to some predefined humidity and temperature levels. However, this approach is often unfeasible or even impossible. Therefore, in this paper we analyzed a dataset of 1\,852 DMS measurements of surgical smoke evaporated from porcine adipose and muscle tissue to get an understanding of the impact of varying humidity and temperature on DMS measurements. Our analysis confirmed clear dependence of the measurements on these two factors. To overcome this challenge, we fitted regression models to raw and normalized DMS measurement data. Subsequently, these models were used for estimating DMS measurements for known tissue types based on recorded humidity and temperatures. Our test suggests that it is possible to estimate DMS measurements of surgical smoke from porcine adipose and muscle tissue under specific environmental conditions by standardizing DMS measurements separation voltage-wise and training multivariate regression models on the normalized data, which is the first step in removing the need for standardized measurement conditions.
\end{abstract}


\section{Introduction}

Ion mobility spectrometry (IMS) is an emerging and promising technique for molecular analysis of complex mixtures of chemicals, which ionizes gas-phase molecules and measures drift times of the resulting ions in a carrier gas under the influence of electric fields. Due to differences in the ions’ molecular weights, charge, and the shape among compounds, their mobilities differ~\cite{Zamora2012}, resulting in different measurements for different volatile organic compounds (VOCs)~\cite{Muller2019}. Therefore, IMS can, unlike mass spectrometers, distinguish molecules with the same mass but different shape. Due to numerous choices of carrier gases and designs for the electric fields, multiple IMS configurations exist~\cite{Dodds2019}. Differential mobility spectrometry (DMS), also known as Field Asymmetric Ion Mobility Spectrometry, operate at atmospheric pressures and yield high degree of selectivity~\cite{Dodds2019}. In DMS an electric field consisting of an asymmetric oscillating high intensity field and a low static field component is used to obtain information on the electric field-mobility dependence of ions~\cite{Anttalainen2019}.

DMS sensors have gained popularity over recent years. DMS measurements of evaporated VOCs as input data for various machine learning techniques has been shown to enable accurate identification of, for example, chemicals~\cite{Rauhameri2024}, diseases in cultivars~\cite{Kothawade2025}, and differentiation of porcine tissue types~\cite{Kontunen2018,Karjalainen2018,Lepomaki2022} as well as distinguishing between cancerous and healthy tissue based on surgical smoke generated from cutting the tissue with a surgical knife or evaporating it with a laser~\cite{Haapala2022,Hermelo2025}.

However, a challenge to the accurate identification based on DMS measurements is that all classifiers require stable environmental conditions. Changes in humidity and temperature impact the mobility of ions~\cite[p.\,250\,ff.]{Eiceman2014}, and hence result in different sensor responses of the same VOCs under different environmental conditions. If stable humidity, temperature, and air pressure cannot be ensured, the performance of any classifier relying on DMS measurements will suffer considerably. A descriptive example of this risk was observed in~\cite{Kontunen2021}. In this study, human tissue samples were identified based on DMS measurements of intraoperatively gathered surgical smoke samples in 20 breast cancer surgeries. The relative humidity varied between 12\% and 37\% over the 20 surgeries, and the linear discriminant analysis classifier, which yielded accuracies between 93 and 95\% in~\cite{Kontunen2018}, only achieved a mean accuracy of 44\%. Other studies also revealed variations in environmental conditions inside operating rooms. For example, in~\cite{Cheng2021} and~\cite{Chen2022} the authors observed temperatures between 19°C and 23°C and relative humidities between 55\% and 65\% over 23 breast surgeries and 22 head and neck surgeries respectively.

Studies in~\cite{Kontunen2021}--\cite{Chen2022} highlighted that regulating humidity and temperature to some predefined standard conditions, in operating rooms, is often unfeasible or even impossible. In this paper, to our knowledge for the first time, the aim was to standardize the measurements rather than the measurement conditions. Therefore, we first investigated the effect of changing humidity and temperature on DMS measurements of surgical smoke from porcine adipose and skeletal muscle tissue. We then tested two regression methods for estimating the DMS measurements of both tissue types under given environmental conditions and compared the estimates with actual DMS measurements at these conditions.

Previous work on predicting DMS measurements include, e.g. ~\cite{Ieritano2021, Iertiano2022}. In~\cite{ Ieritano2021}the authors predicted DMS curves for MeOH and MeCN diluted in water using a random forest regression model and nitrogen as carrier gas. They did not investigate the impact of environmental conditions. In contrast, we used purified, compressed air as carrier gas and predicted DMS measurements of surgical smoke, which is more complex than the diluted chemicals used in~\cite{Ieritano2021}. In~\cite{Iertiano2022} authors described how to predict VOCs DMS measurement by modeling dynamic ion-solvent cluster formation within the DMS sensor. Again, nitrogen was employed as carrier gas and the impact of changing humidity and temperature on DMS measurements was not studied.

The complexity of the ionization chemistry of mixtures presented a major challenge in our work, which was further highlighted by vapor modifier effects (particularly water). This might explain why the earlier works mentioned in the previous paragraph focused on pure chemicals rather than chemical mixtures. The complexity of surgical smoke was shown, for example, in Weber and Splei{\ss}~\cite{Weber1997} who detected main volatile organic substances from eleven chemical groups in surgical smoke from porcine muscle tissue. Albrecht et al.~\cite{Albrecht1994} detected approximately 150 VOCs and identified 15 substances in emission from laser surgery and electrosurgery for liver and fat tissue; Splei{\ss} et al.~\cite{Spleiss1995} identified 59 substances in emissions form surgical CO$_2$ laser of liver and muscle tissue; Barrett and Garber~\cite{Barrett2004} listed 41 chemicals identified within surgical smoke; Al Sahaf et al.~\cite{AlSahaf2007} detected aldehydes, alkanes, and aromatic benzenes in surgical smoke with greater quantities of aldehydes in adipose tissue; Yeganeh et al.~\cite{Yeganeh2020} detected between 12 and 21 VOCs in smoke from smoke from electrocautery of human meniscus, ligament, adipose, muscle, and synovium tissue; and Cheng et al.~\cite{Cheng2021} identified 87 VOCs in surgical smoke samples from breast surgeries. Due to the large number of constituents in surgical smoke, traditional methods for extracting information from DMS measurements of the smoke collected in varying environmental conditions are unlikely to succeed. Hence, in this paper statistical methods are employed to improve our understanding of the obtained DMS measurements.

The contribution of our article is three-fold. First, we demonstrate that even in a controlled laboratory environment, DMS measurements of surgical smoke evaporated from porcine adipose and skeletal muscle tissue vary considerably. Second, we show that these variations are partly due to fluctuations in the observed temperature and absolute humidity. Third, we validate that by standardizing DMS measurements of surgical smoke separation voltage-wise and training multivariate regression models on the normalized data, the impact of changing environmental conditions can be reduced considerably.


\section{Impact of humidity and temperature on Differential Mobility Spectrometry}

Increases in humidity in the internal atmosphere of a DMS analyzer affect both the ionization of chemical substances and the separation of ions derived from these substances, here volatile organic compounds (VOCs). The combined influence of these two effects on overall analytical performance of a DMS analyzer is non-linear yet understandable as separate principles. Sample ionization occurs in positive polarity through gas-phase, ion-molecule reactions between a substance $M$ and commonly hydrated protons $H^{+}(H_2O)_n$, where $n$ is dependent on humidity levels and temperature. Ionization can be understood as a displacement reaction where the electrostatic association of $M$ to a hydrated proton results in water displacement and the formation of a so-called product ion, such as a protonated monomer $MH^{+}(H_2O)_{n-1}$. Although these reactions, at ambient pressure, are best described using reaction enthalpies, proton affinities are an approximate measure of the reactivity of the hydrated proton and various substances. The impact of humidity is seen principally in the hydration number of $H^{+}(H_2O)_n$ where an increase in humidity in isothermal conditions causes an increase in $n$, or the size of the cluster ion. Significantly, the increased size of the cluster causes an increase in the proton affinity of the hydrated proton, and lessened reactivity with substances. Values for proton affinity (kJ/mol) will change with increased cluster size as 691 for $H^{+}(H_2O)$, 911 for $H^{+}(H_2O)_3$, to 1\,040 for $H^{+}(H_2O)_5$~\cite{Safaei2019SciRep,Borsdorf2015}.

Temperature impacts the size of cluster ions: an increase in temperature at fixed humidity results in decreased hydration levels and increased reactivity of the hydrated proton. While VOCs with proton affinities from alcohols to organophosphorus compounds will undergo favorable reactions, or exothermic associations, with clusters of low hydrate numbers at humidites of 1 to 10\,ppm and 50°C, only compounds with proton affinities equal or greater to ketones will be detected at the same temperature above 100\,ppm~\cite{Safaei2019ACA}. In general, only compounds with large proton affinities, such as amines, will be observed clearly in spectra above 10\,000 ppm humidity~\cite{Mäkinen2011}. Nonetheless, responses to compounds of low proton affinity such as benzene and methylated benzenes can be extracted computationally from DMS spectra at humidities up to 77\% relative humidity~\cite{Szczurek2017}.

The separation of ion peaks in DMS spectra are also affected by humidity through processes associated with the mobility of ions under the extremes of electric field strength $E$ in the DMS analyzer. Ions such as protonated monomers described above are pushed by a gas flow between two parallel plates experiencing an oscillating asymmetric electric field between -1\,500\,V/cm to 30\,000\, V/cm (usually 30\% duty cycle at 1 to 5\,MHz). Although the ion core $MH^{+}$ of volatile organic compounds is unchanged by the electric fields, ion clusters are significantly desolvated at 30\,000\,V/cm due to electric field heating. In contrast, when the ion is cooled at low field strengths, re-clustering occurs with water or other small neutrals in the supporting gas atmosphere. Differences in mobility coefficients $K$ between the “cool” highly clustered, and the “hot” declustered forms provide a basis for ion separations according to tendencies to undergo cluster formation and by ion mass. The field dependence of $E$ is monotonic and described as a function, alpha $\alpha (E/N)$, taken from $K_0(E/N) = K (1 + \alpha (E/N) )$, where $K_0$ is a field dependent reduced mobility coefficient and $E/N$ is the field strength normalized to gas number density $N$. This is seen in DMS spectra as the position of a peak in the compensation voltage (CV) scale. Ions with large alpha functions are found at increased distances (dispersed) from the reference point (0\,V CV) where ions without any mobility dependence on $E/N$ are located. 

An increase in humidity (or other small neutrals) and the associated increased hydration (or solvation) of the core ion during the low field portion of the waveform causes an increase in the position or slope of the alpha function leading to increased discrimination, i.e., peak separation, of ions. Consequently, increases in humidity of the supporting atmosphere will increase the resolving power of the DMS analyzer and the resolution among peaks in DMS spectra. In practice, clusters form when there is enough time for ion collisions with neutrals during the low-field portion of a waveform ($\approx$100\,ns). In practice ion-neutral collisions are observed when humidity is 50\,ppm and greater. Thus, there is no influence on ion peak positions, i.e., CV voltage in DMS spectra, at humiditys below this limit. The impact on peak position at values greater than 50\,ppm however is large. For example, Krylova et al.~\cite{Krylova2003} observed that increasing humidity from 1\,000 to 10\,000 ppm resulted in a two-fold or higher increase in the alpha plot with increased dispersion in the CV scale. At extreme levels of humidity, say 10\,000\,ppm and higher, the hydration level of an ion is so high that ion desolvation at high $E/N$, and ion dispersion in the CV axis, is negligible. The electric field will be unable to change the hydration levels since the ions will be 'too' wet. For example, Kuklya et al.~\cite{Kuklya2015} observed that the reactant ion peak, which is the peak related to the lowest proton affinities, trended towards increased CV values as the humidity was increased from 6.5 to 20\,ppm$_{\text{v}}$; however, as the humidity was increased to 550\,ppm$_{\text{v}}$ the alpha function flattened and the ion peak CV trended in the direction of 0\,V.

A second experimental variable in DMS measurements that affects the quantitative response and CV values for peaks in DMS spectra is temperature. Since the CV value for peaks of small ions in DMS spectra is governed by differences in ion cluster size at field extremes, increased temperature of the supporting atmosphere will decluster ions and reduce the differences. When ions at a low field are strongly heated by gas temperature, they exhibit similar mobility behavior as the ions already dehydrated at large $E/N$. While hand-held DMS analyzers are operated at ambient temperature, laboratory grade analyzers are heated. The increased temperature can be used to decluster hydrated protons, providing improved reactivity and improved peak dispersion on the CV scale, over unheated analyzers. Although the influence of temperature on ion hydration has been described for halide ions in low field ion mobility spectrometry~\cite{Wolanska2023}, the three variables of humidity levels, temperature, and waveform fields have not yet been formalized with an integrated practice for response and ion characterization to a range of small to medium size volatile organic compounds with DMS.

Krylov et al.~\cite{Krylov2009} studied the impact of temperatures varying between 25 and 150°C on DMS measurements of three positive ions (methyl salicylate, dimethyl methylphosphonate, and toluene), and one negative ion (methyl salicylatande). They detected peak values for separation voltages ranging from 500 to 1\,500V and observed strong variations in the compensation voltages at which the highest responses were observed. They concluded that the locations of alpha curves were depending considerably on the temperature for all four investigated ions. While the processes of cluster formation and dissociation occur for all ions in DMS measurements, effects on alpha functions (or differences in $K_0$ at extreme field strengths) are negligible or even nil on large ions (e.g., >500\,Da) as the addition of one or a few waters does not cause a measurable change in ion motion.


\section{Data and Methods}

\subsection{Data Collection and Description}

A total of 1\,852 DMS measurements from surgical smoke were collected. The tissue samples (1\,089 adipose, 763 muscle tissue) were gathered from nine healthy Finnish landrace pigs (\textit{Sus scrofa domesticus}) and placed in laser matrices. The automated tissue laser analysis system for imaging approaches (iATLAS) described in~\cite{Lepomaki2022} and reference therein was used for sampling, which utilizes an ENVI-AMC DMS unit (Environics Oy, Finland) coupled with a laser evaporation unit. More details on the data collection procedure can be found in~\cite{Lepomaki2022}.

DMS measurements are commonly represented by image-like dispersion plots. In these plots, the x-axis represents the compensation voltage, while the y-axis represents the separation voltage (SV), and the z-axis (i.e. the pixel value) represents the response for the corresponding CV/SV combination, measured in pA. In this study, the CV sweep was set from -0.8\,V to 5\,V with 40 steps and the SV sweep was set from 350\,V to 700\,V with 12 steps, meaning that the dispersion plots were representations of 12-by-40 matrices. Only dispersion plots for positive ions were considered for further analysis.

\begin{figure}
\centering
\includegraphics[width=.98\columnwidth,clip=true, trim=2cm .8cm 4cm 2cm]
{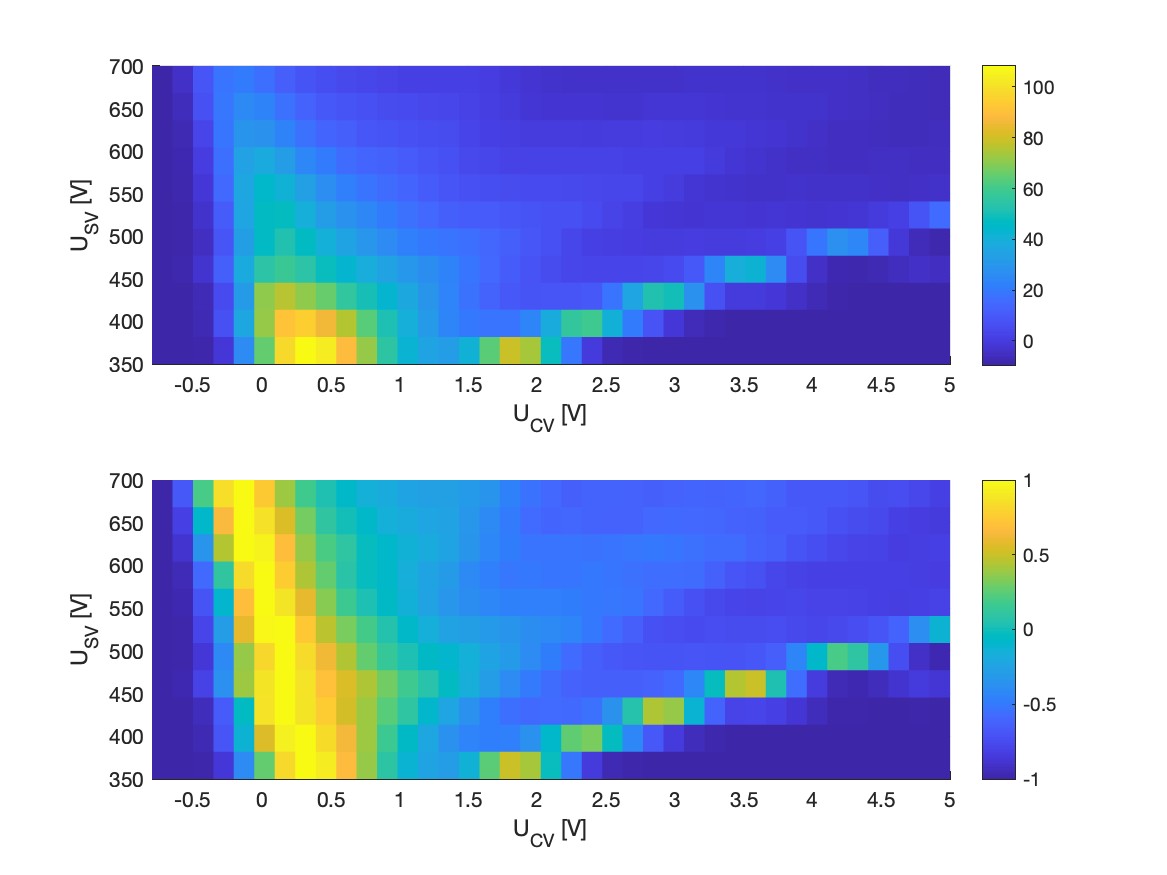}
\caption{
Example of a dispersion plot for surgical smoke from adipose tissue (top) and dispersion plot with data row-wise standardized to $[-1,1]$ from the same sample (bottom).}
\label{fig:exDP}
\vspace{-2ex}
\end{figure}

A typical dispersion plot from DMS analysis of vapors from laser cutting of adipose tissue is shown in Figure~\ref{fig:exDP} for the original (top) and processed (bottom) data from the same sample. The patterns barely visible in the plot for original data are accentuated after row-wise normalization to $[-1,1]$ (bottom), which means that all responses for a fixed SV are standardized. This approach was suggested in~\cite{ Rauhameri2024,Virtanen2022}, although the latter one employed SV-wise normalization to $[0,1]$. After data normalization several features can be seen for ion peak when measured over the range of separation voltages. A dispersion pattern is seen between CVs of 1.6 and 5\,V, which is characteristic of a polar ion of small mass such as a hydrated proton, $H^{+}(H_2O)_n$, as described above. Alternatively, the intensity and persistence of this ion peak are also consistent with hydrated ammonium, $NH_4^{+}(H_2O)_n$, also a small polar ion with high proton affinity. Both of these ions are commonly seen in DMS measurements with purified air atmospheres and the slope of the dispersion plots suggests strong, positive alpha coefficients known for these ions. The dispersion pattern of these hydrated protons, which in general are used for ionization of the analyte, are often called the reactant ion peak when referring to DMS spectra.

Another feature of this dispersion plot is the intense ion peak whose plot extends in CV from 0.4 to -0.3\,V. Such plots are generally referred to as alpha curves in the context of DMS spectra. The slope of this alpha curve is characteristic of a large ion or an ion which is highly clustered with neutrals, i.e., an ion of comparatively high mass and a negative alpha function. Negative alpha functions commonly occur with ion masses over 150 to 200 Da. An interpretation of this ion, mindful of the anticipated chemical complexity of vapors from this sample, is as an ion cluster which arises where a substance of strong ionization properties is electrostatically bound to unionized constituents in a complex vapor sample. While this condition has historically limited conclusions about the chemical constituents responsible for the ion peak, the pattern may still contain extractable chemical information. Significantly, there are some profiles or features which can be seen between the extremes of these two features. These are weakly visible above the baseline at intensities of -0.3 to 0.4\,V. Although these are difficult to evaluated visually, computational methods may extract chemical information from such patterns. In summary, the dispersion plots are consistent with a complex chemical mixture, as anticipated from vapors generated from adipose tissue. More details on dispersion plots and their descriptive sections can be found in~\cite{Anttalainen2019}.

\subsection{Analysis of environmental conditions}

Although the measurements were collected in a controlled laboratory environment, recorded temperature of the sample $t_{\text{meas}}$, the internal temperature measured within the field effect transistor (FET)-based temperature sensor $t_{\text{FET}}$ of the DMS device, and absolute humidity $h_{\text{abs}}$ varied considerably during data collection, as can be seen in Table~\ref{tab:envCond}.
%
\begin{table}
\renewcommand{\arraystretch}{1.3}
\caption{{\rm Minima, 2.5\%-quantiles, median values, 97.5\%-quantiles, and maxima for $t_{\text{meas}}$, $t_{\text{FET}}$, and $h_{\text{abs}}$during the measurement campaign.
}}
\label{tab:envCond}
\centering
\begin{tabular}{llllcc}
factor & min. & 2.5\%-qu. & median & 97.5\%-qu. & max. \\ [0.5ex]
\hline
$t_{\text{meas}}$ & 24.7°C & 28.2°C  & 30.0°C & 31.0°C & 31.1°C \\
$t_{\text{FET}}$ & 41.9°C & 44.2°C & 45.7°C & 48.0°C & 48.2°C \\
$h_{\text{abs}}$ & 1.84$\frac{\text{g}}{\text{m}^3}$ & 2.09$\frac{\text{g}}{\text{m}^3}$ & 2.48$\frac{\text{g}}{\text{m}^3}$ & 2.63$\frac{\text{g}}{\text{m}^3}$ & 2.66$\frac{\text{g}}{\text{m}^3}$ 
\end{tabular}
\end{table}

\begin{figure}
\centering
\includegraphics[width=.98\columnwidth, clip=true, trim=2.2cm 1cm 2.9cm 1cm]{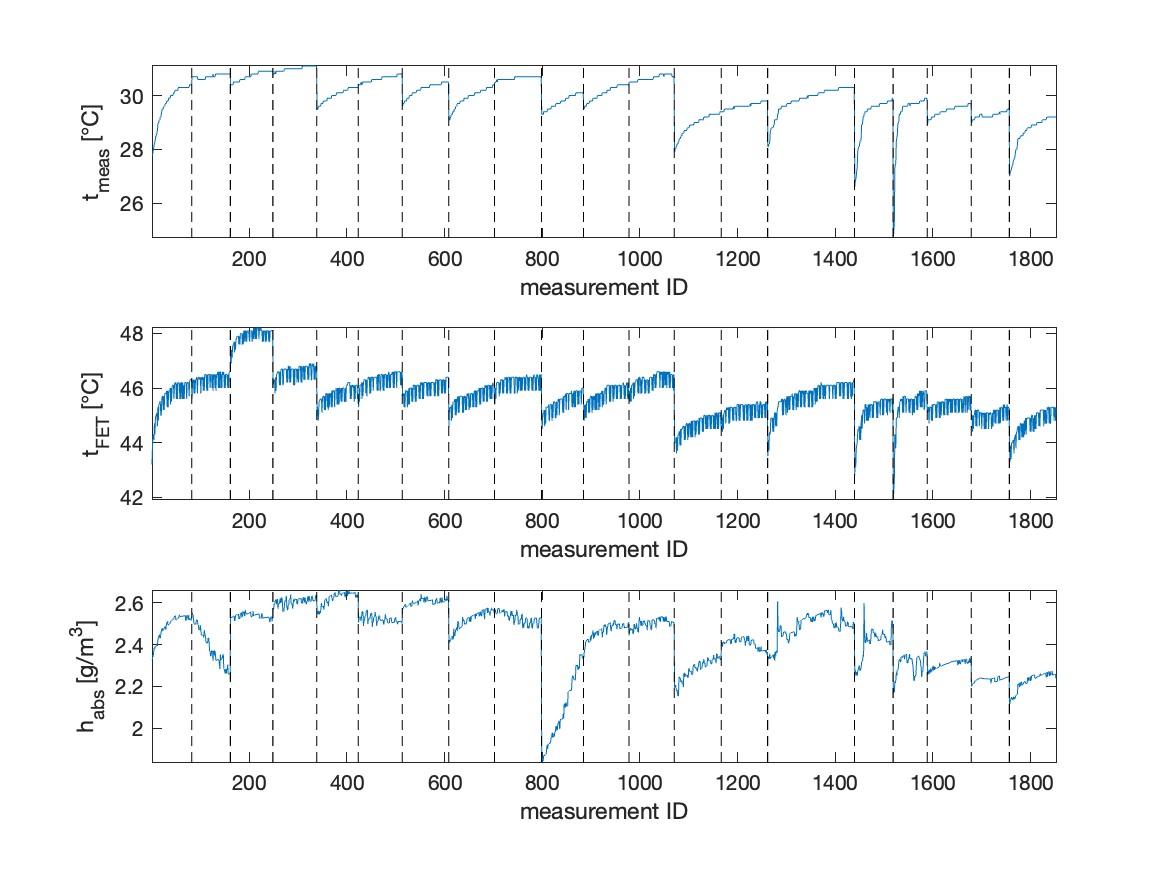}
\caption{
Measurement temperature (top), FET temperature (middle), and absolute humidity (bottom) over all 1\,852 measurements ordered according to their IDs.
}
\label{fig:Temp}
\vspace{-2ex}
\end{figure}

Figure~\ref{fig:Temp} shows in the upper plot the measurement temperatures recorded for all 1\,852 measurements sorted according to their IDs, i.e. with respect to the order in which the samples were collected. The figure shows a repetitive pattern where $t_{\text{meas}}$ increases approximately logarithmically, due to the measurement system warming up during its operation, before a sudden drop between two consecutive measurements. These rapid changes occurred at switches from one sample matrix to another. The switches are indicated by vertical dashed lines in the three plots of Figure~\ref{fig:Temp}. Switching between sample matrices lasted always a few minutes during which the sampling system cooled down. Furthermore, data was collected over multiple sessions distributed over 9 days, which explains the majority of the 19 switches marked in the figure. The FET temperatures showed a similar behavior as the measurement temperatures over all measurements sorted with respect to the sampling order, although with an average offset of 15.81°C and stronger fluctuations between subsequent samples (compare the center plot in Figure~\ref{fig:Temp}). Both temperatures were highly linearly correlated with a Pearson correlation of 0.85 ($p$-value = 0 within machine precision; Spearman correlation coefficient of 0.90 with $p$-value = 0).

\begin{figure}
\centering
\includegraphics[width=.98\columnwidth,clip=true, trim=2.8cm 1cm 2.9cm 1cm]{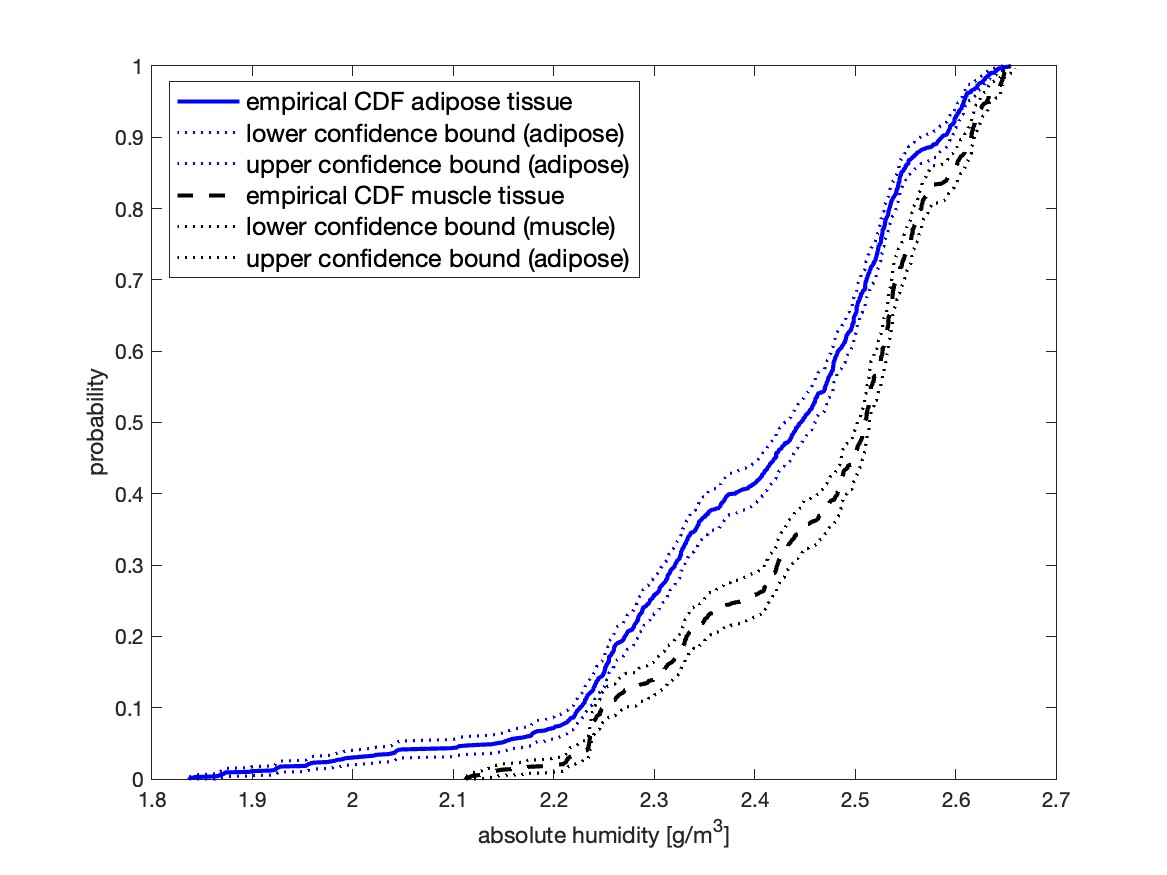}
\caption{
Empirical cumulative distribution functions of recorded absolute humidities for adipose (solid, blue line) and muscle tissue samples (dashed, black line) with dotted lines indicates boundaries of 95\% confidence intervals.
}
\label{fig:CDFs}
\vspace{-2ex}
\end{figure}

Absolute humidity readings were moderately linearly correlated to $t_{\text{meas}}$ (0.6506 with $p$-value = 2.35$\cdot 10^{-223}$) and $t_{\text{FET}}$ (0.5839 with $p$-value = 1.06 $\cdot 10^{-169}$). Spearman correlation coefficients were slightly higher: 0.70 with $t_{\text{meas}}$ ($p$-value = 7.70 $\cdot 10^{-273}$) and 0.67 $t_{\text{FET}}$ ($p$-value = 1.22 $\cdot 10^{-237}$). The bottom plot in Figure~\ref{fig:Temp} shows upward and downward movements for absolute humidity, with jumps at some of the switches between sample matrices. Since the behavior differs from that of the measurement temperature, other factors most likely contributed to the observed changes. One source of humidity are the tissue samples themselves. For example, lean muscle tissue consists of approximately 75\% water~\cite{Huff2005}. The burning of the tissue samples with the laser system evaporates, amongst other components, water from the cells. Hence, increases in the absolute humidity can be expected. Contrary, adipose tissue is mainly composed of adipocytes and its water content is considerably lower. For example, Wang and Pierson~\cite{Wang1976} reported an average water content of 14.4\% in adipose tissue from surgical biopsy samples from 16 patients undergoing elective laparotomy. Thomas~\cite{Thomas1962} analyzed samples from 59 surgeries and observed average water contents between 9.6\% and 14.4\%, noticing an increase in water content the leaner the patient was. Thus, burning adipose tissue should result in smaller absolute humidity levels measured by the DMS device. This hypothesis is supported by the empirical cumulative distribution functions (CDFs) of the recorded absolute humidity levels for all 1\,089 adipose and all 763 muscle tissue samples (see Figure~\ref{fig:CDFs}). For any probability level, the corresponding absolute humidity for muscle tissue samples is higher than the absolute humidity for adipose tissue samples. Even the 95\% confidence interval boundaries of both CDFs do not overlap except for probabilities above 0.99. Therefore, changes in temperatures and absolute humidity were, at least partly, caused by the sampling system and could not be prevented without unreasonable hardware modifications or changes to the measuring protocol (e.g. waiting after every DMS measurement for the temperature to return to the starting temperature).

Based on these separate observations for the impacts of humidity and temperature on DMS measurements (see Section II for details) it is clear that combined changes of humidity and temperature also alter DMS measurements. An attempt to model the impact of changing humidity and temperature was published in~\cite{Wolanska2023}. Authors developed a theoretical model for calculating the effective mobility of ions at given humidity and temperature. The model assumed that effective mobility coefficients depended linearly on the mobility of ions at medium- and high-water vapor concentrations in the drift gas. In~\cite{Wolanska2023}, only the effect of humidity on small ions (hydronium, ammonium, oxygen, chloride, bromide, and iodide ions) at various temperatures was studied. Hence, the theoretical model from~\cite{Wolanska2023} might be overly simplistic and not directly applicable for highly complex mixtures such as surgical smoke.

\begin{figure}
\centering
\includegraphics[width=.98\columnwidth,clip=true, trim=2cm .8cm 4cm 2cm]
{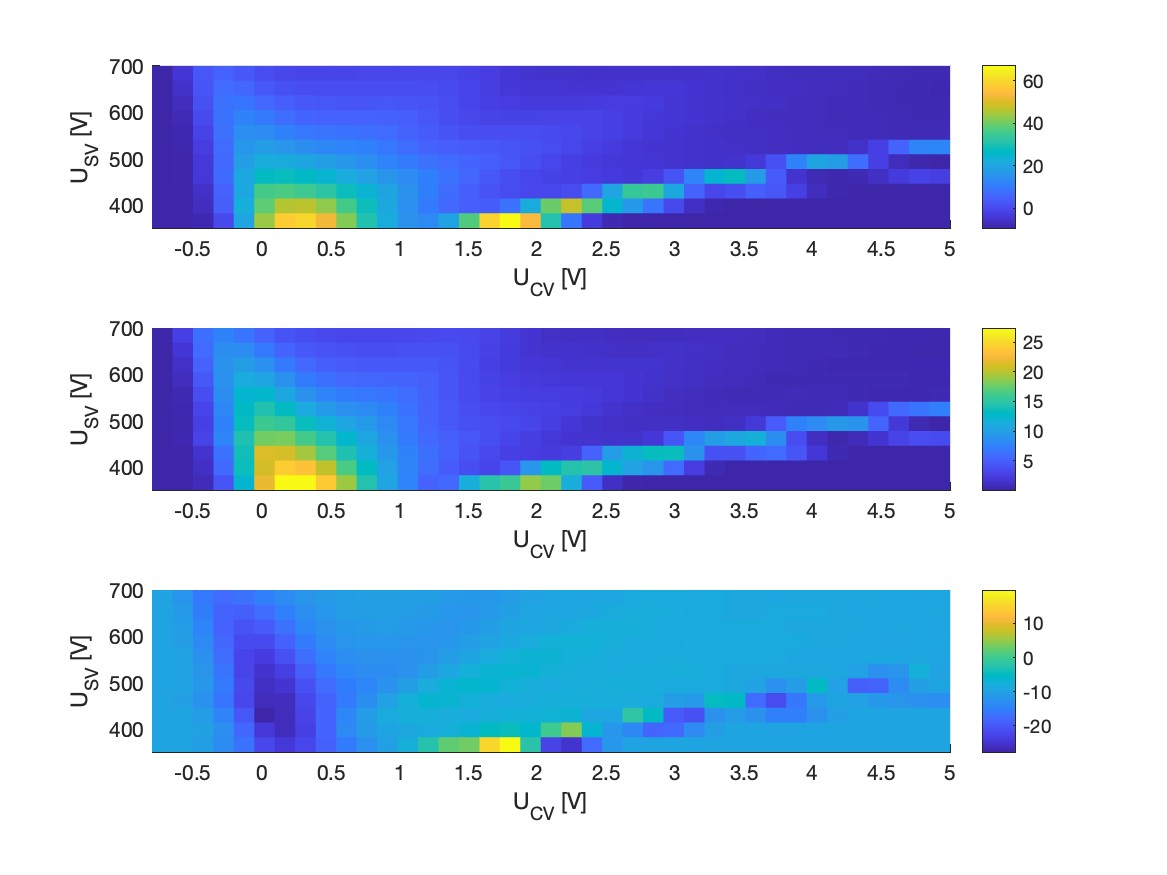}
\caption{
Average dispersion plot for adipose tissue (top) and corresponding CV/SV-wise standard deviations (middle), as well as the differences between the average responses and threefold standard deviations (bottom). Responses in pA.}
\label{fig:avg+sdAdi}
\vspace{-2ex}
\end{figure}

\begin{figure}
\centering
\includegraphics[width=.98\columnwidth,clip=true, trim=2cm .8cm 4cm 2cm]
{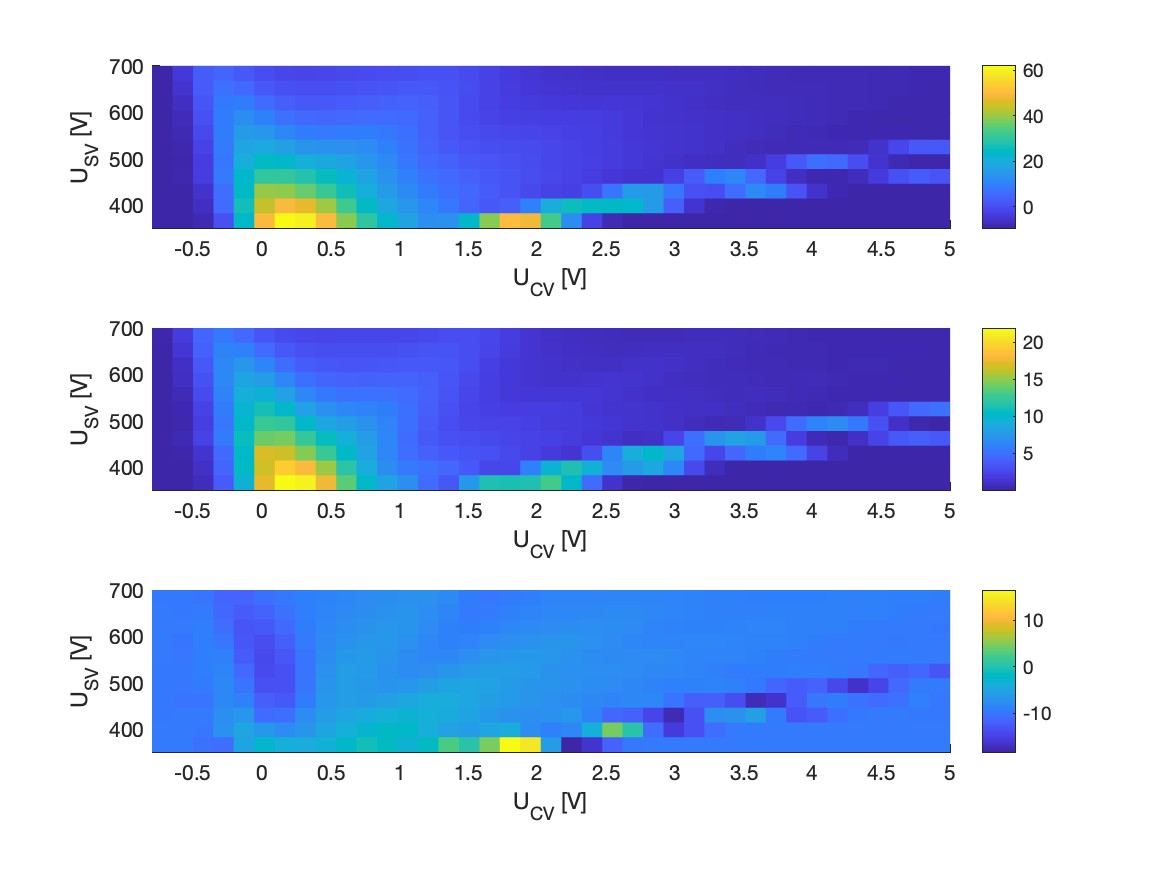}
\caption{
Average dispersion plot for muscle tissue (top) and corresponding CV/SV-wise standard deviations (middle), as well as the differences between the average responses and threefold standard deviations (bottom). Responses in pA.}
\label{fig:avg+sdMus}
\vspace{-2ex}
\end{figure}

Figures~\ref{fig:avg+sdAdi} and~\ref{fig:avg+sdMus} illustrate the impact that changes in temperature and humidity had on the DMS measurements analyzed in this manuscript. Both figures show the average dispersion plots, their corresponding standard deviations, as well as the differences between the average dispersion plots and the 3$\sigma$ dispersion plots for adipose and muscle samples respectively. In the average plots, regions with average responses well above zero indicate the alpha curves caused by ion peaks that are well above the noise level~\cite{Anttalainen2019}. These alpha curves are the regions of interest when interpreting dispersion plots and are used for tissue identification based on DMS measurements. The corresponding standard deviation plots in Figures~\ref{fig:avg+sdAdi} and~\ref{fig:avg+sdMus} reveal that the locations of the alpha curves varied considerably. The differences between the average responses and threefold the standard deviation of responses are close to zero or even negative for a large part of the locations of branches. To conclude, not only the strength of the ion peaks changed but also the locations of the branches caused by ion peaks shifted between DMS measurements of the same tissue type. Hence, methods for compensating or, at least, mitigating the impact of changing environmental conditions on DMS measurements should be developed.


\subsection{Regression Methods and Analysis}

In order to decide which regression models to test for mitigating the impact of temperature and humidity, currents of the DMS device to specific combinations of CV and SV for all smoke sample measurements from either adipose or muscle tissue were sorted in ascending order with respect to a) $h_{\text{abs}}$, b) $t_{\text{meas}}$, and c) $t_{\text{FET}}$ . Afterwards, linear functions of the form $c = a*x+b$ were fitted to the sorted measurements, where $x$ was the running sample number, $a$ was the slope, and $b$ the intercept. Table~\ref{tab:linmdlcoeff} summarizes median values, as well as 2.5\%- and 97.5\%-quantiles for the slope coefficients $a$. The intercepts $b$ are merely offsets that provide no information on trends in the sequences and, hence, are omitted. Coefficient $a$ is always approximately zero, independent on the tissue type and the sorting criteria, which indicates that there is no clear upward or downward trend in the measured current for increasing $h_{\text{abs}}$, $t_{\text{meas}}$, or $t_{\text{FET}}$.
%
\begin{table}
\renewcommand{\arraystretch}{1.3}
\caption{{\rm Statistics for slopes of linear functions fitted to currents measured for specific combinations of CV and SV. For all smoke sample measurements from either adipose or muscle tissue linear functions where fitted to currents sorted in ascending order with respect to $h_{\text{abs}}$, $t_{\text{meas}}$ as well as $t_{\text{FET}}$.
}}
\label{tab:linmdlcoeff}
\centering
\begin{tabular}{lccccc}
tissue & coeff. & factor &  2.5\%-qu. & median & 97.5\%-qu. \\ [0.5ex]
\hline
adipose & $a$ & $h_{\text{abs}}$ & -0.0046 & 0.0014 & 0.0136 \\
 & $a$ & $t_{\text{meas}}$ & -0.0005 & 0.0015 & 0.0215 \\
 & $a$ & $t_{\text{FET}}$ & -0.0005 & 0.0018  & 0.0205 \\
 muscle & $a$ & $h_{\text{abs}}$ & -0.0059 & 0.0021 & 0.0226 \\
 & $a$ & $t_{\text{meas}}$ & -0.0067 & 0.0011 & 0.0219 \\
 & $a$ & $t_{\text{FET}}$ & -0.0066 & 0.0010  & 0.0218
\end{tabular}
\end{table}
%

\begin{figure}
\centering
\includegraphics[width=.98\columnwidth,clip=true, trim=2.7cm 1cm 2.9cm 1cm]{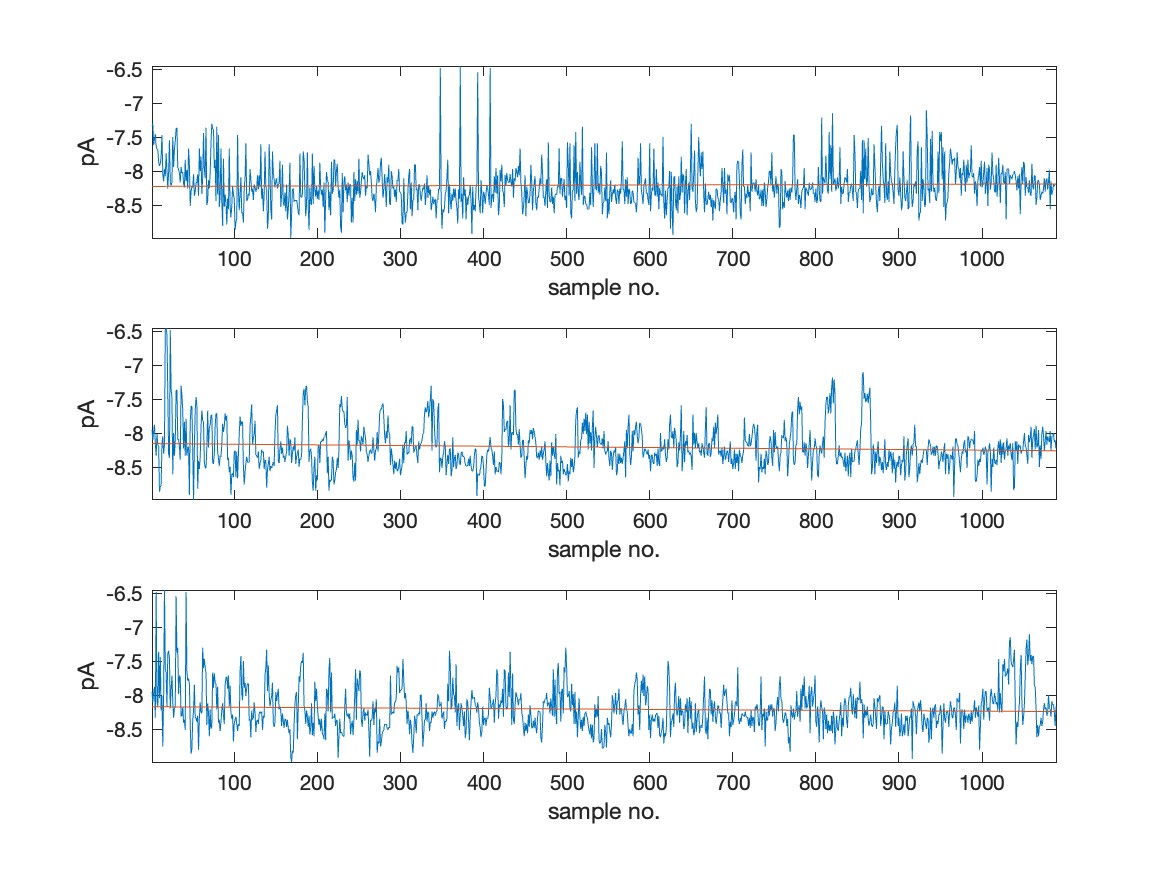}
\caption{
Currents measured over all 1\,089 adipose smoke samples for CV=4.855\,V and SV=612.500\,V sorted with respect to $h_{\text{abs}}$ (top), $t_{\text{FET}}$ (middle), and $t_{\text{meas}}$ (bottom).
}
\label{fig:linMdlDecAdipose}
\vspace{-2ex}
\end{figure}

\begin{figure}
\centering
\includegraphics[width=.98\columnwidth,clip=true, trim=3cm 1cm 2.9cm 1cm]{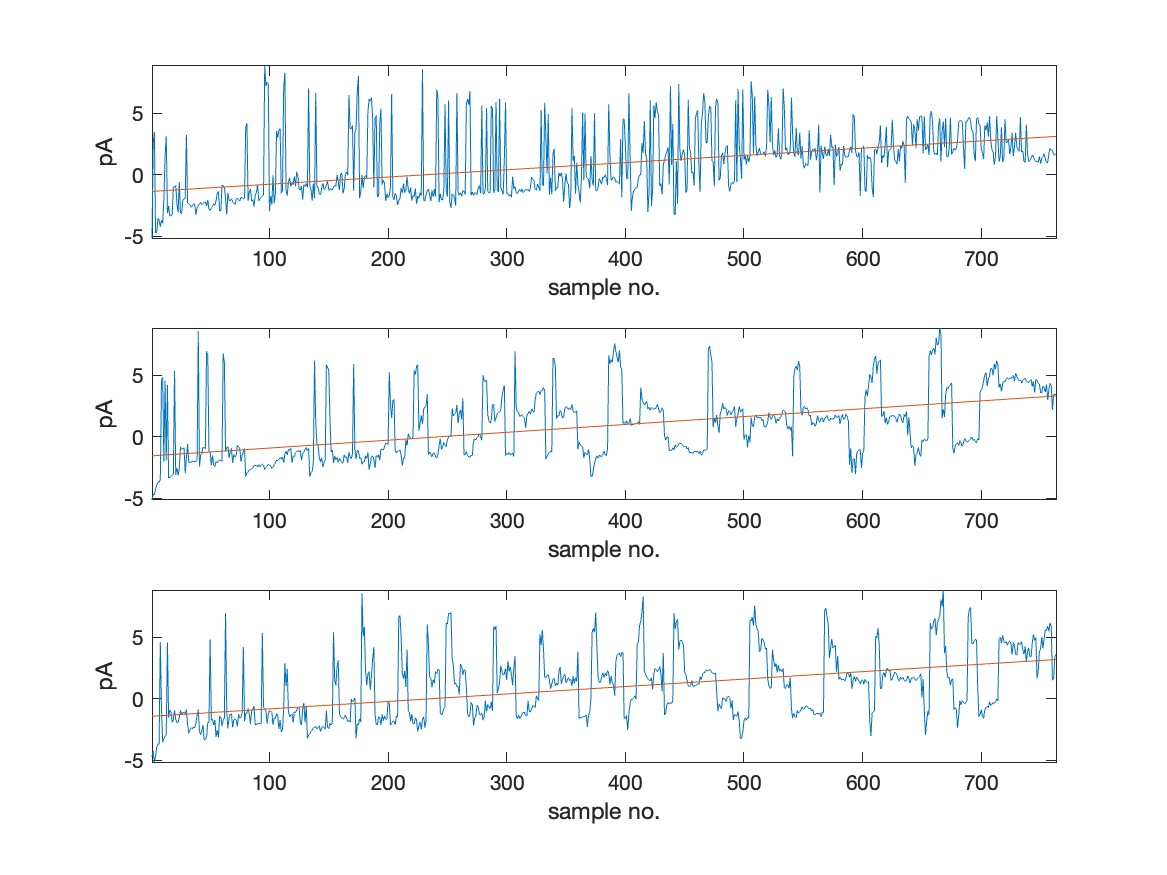}
\caption{
Currents measured over all 763 muscle smoke samples for CV=1.085\,V and SV=641.667\,V sorted with respect to $h_{\text{abs}}$ (top), $t_{\text{FET}}$ (middle), and $t_{\text{meas}}$ (bottom).
}
\label{fig:linMdlDecMuscle}
\vspace{-2ex}
\end{figure}

Figures~\ref{fig:linMdlDecAdipose} and~\ref{fig:linMdlDecMuscle} show examples of sorted currents with fitted linear functions for two specific CV-SV combinations. The plots illustrate the typical behavior of sorted currents with fitted linear functions. No repetitive pattern could be detected even within a single sorted set of measured currents. Furthermore, all three environmental parameters $h_{\text{abs}}$, $t_{\text{meas}}$ and $t_{\text{FET}}$ were believed to influence the measured currents. Thus, fitting any high-order polynomial seemed unpromising. Hence, in this paper multiple linear regression and multivariate regression were tested for predicting DMS measurements at given absolute humidity, sample temperature and the internal temperature measured within the DMS device's FET-based temperature sensor.


The tested linear regression model estimated the response $y_{\text{CV,SV}}$ for a given CV/SV combination using
\begin{equation}
y_{\text{CV,SV}} = a_0 + a_1 t_{\text{meas}} + a_2 t_{\text{FET}} + a_3 h_{\text{abs}},
\end{equation}
where $a_0$, $a_1$, $a_2$, and $a_3$ were the coefficients to be estimated from responses for the CV/SV combination for which $t_{\text{meas}}$, $t_{\text{FET}}$, and $h_{\text{abs}}$ were known. This means, each pixel in the dispersion plots was estimated by a different regression model. The linear regression model was chosen due to its simplicity and effectiveness for approximately linear relationships between the scalar input ($t_{\text{meas}}$, $t_{\text{FET}}$, and $h_{\text{abs}}$) and scalar output ($y_{\text{CV,SV}}$) variables. Although linear regression models are known to be sensitive to outliers, this was not considered an issue based on our preliminary analysis of the dispersion plots and the large number of samples available.

In contrast, the tested multivariate regression model estimated the $n$-dimensional response vector ${\bf{y}}_{\text{SV}} = \{ y_{\text{CV}_1,\text{SV}}, .., y_{\text{CV}_n,\text{SV}} \}$ for a given SV using
\begin{equation}
{\bf{y}}_{\text{SV}} = [t_{\text{meas}} \, t_{\text{FET}} \, h_{\text{abs}}] {\bf B} + {\bf e},
\end{equation}
where {\bf B} was a 3-by-$n$ matrix containing the regression coefficients linked to $t_{\text{meas}}$, $t_{\text{FET}}$, and $h_{\text{abs}}$ for each of the $n$ compensation voltages. This means, although the model estimated rows of dispersion plots, each pixel still had its own coefficients. The $n$-dimensional vector {\bf e} contained error terms and followed a multivariate Gaussian distribution, i.e. ${\bf e} ~ \text{MVN} ( {\bf 0}, \Sigma )$ with $n$-dimensional zero vector {\bf 0} and $n$-by-$n$ covariance matrix $\Sigma$. The advantage of multivariate regression models is that they predict, unlike linear regression models, multiple (possibly) correlated scalar responses (${\bf{y}}_{\text{SV}}$). In~\cite{Rauhameri2024}, we demonstrated that interpreting DMS measurements at varying compensation voltages with fixed separation voltages as sequential data is beneficial for classifying chemicals. Sequential data often exhibits correlation. Therefore, we decided to test regression models that would enable capturing potential correlations between the DMS responses for varying compensation voltages with fixed separation voltage.

Both regression models were tested with raw dispersion plot data as well as with row-wise normalized data. Row-wise (i.e. SV-wise) normalization of dispersion plots was first proposed in~\cite{Virtanen2022} for highlighting responses in dispersion plot regions containing alpha curves. Comparing the raw and row-wise normalized plots in Figure~\ref{fig:exDP} illustrates how the approach highlights minor constituents found at high separation voltages while suppressing major constituents found at low separation voltages. The usefulness of row-wise normalization was demonstrated in~\cite{Rauhameri2024} when interpreting dispersion plots as a set of $m$ $n$-dimensional sequential measurements and using long short-term memory neural networks for classification of chemicals based on their DMS spectra. While Virtanen et al.~\cite{Virtanen2022} scaled data to values between 0 and 1 using minimum and maximum responses of the row, we scaled data, similarly as in~\cite{Rauhameri2024}, to values between -1 and 1 to prevent numerical problems when computing the coefficients of the multivariate regression model. Thus, the normalized value $y_{\text{CV},\text{SV}}^{\text{s}}$ for a CV/SV combination was defined as
\begin{equation}
\label{eq:standeq}
y_{\text{CV},\text{SV}}^{\text{s}} = 2 \frac{y_{\text{CV},\text{SV}}  - \text{min}({\bf{y}}_{\text{SV}})}{\text{max}({\bf{y}}_{\text{SV}})  - \text{min}({\bf{y}}_{\text{SV}})} - 1,
\end{equation}
where $\text{min}({\bf{y}}_{\text{SV}})$ and $\text{max}({\bf{y}}_{\text{SV}})$ were the minimum and maximum responses over all compensation voltages at fixed SV.

For testing the estimation performance of the regression methods, a test setup based on 10-fold cross-validation was used. The dataset was divided into two sets containing only DMS measurements from adipose or muscle tissue samples respectively. Each of these two sets were split randomly into ten equally sized subsets, containing each 109 and 76 samples for adipose and muscle data respectively. Next, nine of the ten subsets were used for estimating the coefficients of the regression models. Afterwards, the regression models were used for estimating the DMS measurements in the tenth subset, given their corresponding $t_{\text{meas}}$, $t_{\text{FET}}$, and $h_{\text{abs}}$. In case of training the regression models with normalized DMS data, the de-normalized DMS measurement was obtained by solving (\ref{eq:standeq}) for $y_{\text{CV},\text{SV}}$, namely
\begin{equation}
\label{eq:destandeq}
y_{\text{CV},\text{SV}} = \frac{1}{2} (y_{\text{CV},\text{SV}}^{\text{e}} + 1) ( \text{max}({\bf{y}}_{\text{SV}})  - \text{min}({\bf{y}}_{\text{SV}}) ) + \text{min}({\bf{y}}_{\text{SV}}),
\end{equation}
with $y_{\text{CV},\text{SV}}^{\text{e}}$ being the estimated normalized response. Finally, the differences between estimated and measured DMS measurements were calculated. This was repeated ten times so that each subset was used nine times for training the coefficients of the regression model and once for testing the estimation performance. Thus, the performance of both regression models trained with raw as well as normalized DMS data was evaluated for 1\,090 adipose and for 760 muscle smoke samples. For each trained regression model the coefficient of determination $R^2$, providing a measure for how well the regression model replicated the measured DMS response, was calculated. $R^2$ was defined as
\begin{equation}
\label{eq:R2}
R^2 = 1 - \frac{\sum_{i=1}^{n} (y_{\text{CV},\text{SV},i} - y_{\text{CV},\text{SV},i}^{\text{s}} )^2}{ \sum_{i=1}^{n} (y_{\text{CV},\text{SV},i} - \frac{1}{n} \sum_{i=1}^{n} y_{\text{CV},\text{SV},i})}
\end{equation}
where $n$ was the number of samples used for training the regression model. In addition, for each regression model type CV/SV-wise root mean square errors (RMSEs) were calculated from the differences between estimated and measured DMS measurements of all ten repetitions.

All code was implemented in MATLAB R2023b (MathWorks, USA), using built-in functions \verb+fitlm+ for the linear regression model and \verb+mvregress+ for the multivariate regression model. 


\section{Results and Discussion}

Table~\ref{tab:R2summary} presents averages, standard deviations, medians, and quantiles over $R^2$ values from all trained models for a specific tissue, regression, and data type. Both regression models yielded on average high $R^2$ values with raw as well as normalized data, indicating that the models accounted for most of the variability of the dependent variable in the data sets. However, the multivariate regression models fitted to normalized data yielded the highest averages, median and quantile values for $R^2$ for both adipose and muscle tissue apart from the 97.5\%-quantile for adipose tissue samples. Furthermore, the standard deviation of $R^2$ values was the smallest for multivariate regression models fitted to normalized data. The benefit of these lower standard deviations is especially visible in the 2.5\%-quantiles. Multivariate regression models fitted to the normalized data explain at least 69.91\% (adipose) and 77.79\% (muscle) of the variability in at least 97.5\% of the predicted dispersion plots, while the remaining three methods explained less than 57.43\% (adipose) and 69.35\% (muscle) of the variability in at least 97.5\% of the predicted dispersion plots. The reason for observing higher $R^2$ values for muscle than for adipose tissue is unknown.

\begin{table}[!ht]
\renewcommand{\arraystretch}{1.3}
\caption{{\rm Summary of $R^2$ values for all trained regression models. For each tissue type the highest average, medium, and quantile values of $R^2$ as well as the lowest standard deviations over all four tested regression models are marked in bold.
}}
\label{tab:R2summary}
\centering
\begin{tabular}{@{}lllccccccc@{}}
tissue & regression & data & mean & SD & 2.5\%- & 25\%- & median & 75\%- & 97.5\%- \\
type & type & type & & & qu. & qu. &  & qu. & qu. \\ [0.5ex]
\hline
adipose & linear & raw & 0.8606 & 0.1288 & 0.5564 & 0.8226 & 0.9029 & 0.9465 & \textbf{0.9818}\\
& & norm. & 0.8596 & 0.1213 & 0.5743 & 0.8213 & 0.8989 & 0.9443 & 0.9763\\
& multi- & raw & 0.8463 & 0.1285 & 0.5742 & 0.7991 & 0.8872 & 0.9393 & 0.9766\\
& variate & norm. & \textbf{0.9020} & \textbf{0.0894} & \textbf{0.6991} & \textbf{0.8887} & \textbf{0.9382} & \textbf{0.9575} & 0.9731\\ [0.5ex]
muscle & linear & raw & 0.8978 & 0.1101 & 0.6643 & 0.8601 & 0.9447 & 0.9705 & 0.9892\\
& & norm & 0.8391 & 0.1332 & 0.6394 & 0.7400 & 0.8674 & 0.9562 & 0.9855\\
& multi- & raw & 0.8920 & 0.0973 & 0.6935 & 0.8394 & 0.9239 & 0.9687 & 0.9902\\
& variate & norm & \textbf{0.9393} & \textbf{0.0695} & \textbf{0.7790} & \textbf{0.9337} & \textbf{0.9645} & \textbf{0.9791} & \textbf{0.9920} \\
\end{tabular}
\end{table}

\begin{figure}
\centering
\includegraphics[width=.98\columnwidth,clip=true, trim=11cm .7cm 10.5cm 1cm]{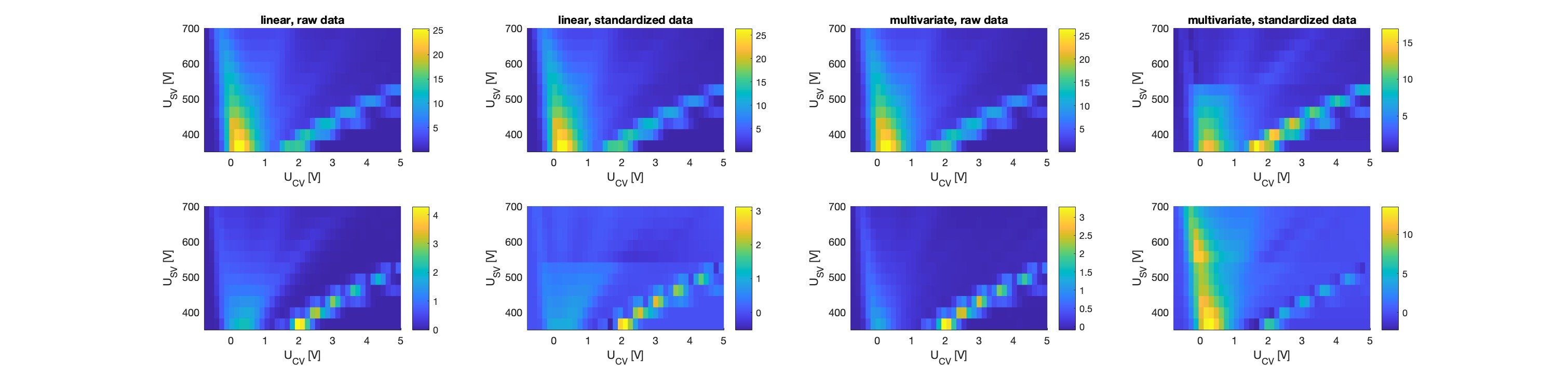}
\caption{
Pixelwise RMSEs for linear and multivariate regression models with raw as well as standardized DMS data from adipose tissue (upper row) and corresponding differences of pixelwise standard deviations and pixelwise RMSEs (lower row). Responses in pA.
}
\label{fig:AdiRes}
\vspace{-2ex}
\end{figure}

\begin{figure}
\centering
\includegraphics[width=.98\columnwidth,clip=true, trim=11cm .7cm 10.5cm 1cm]{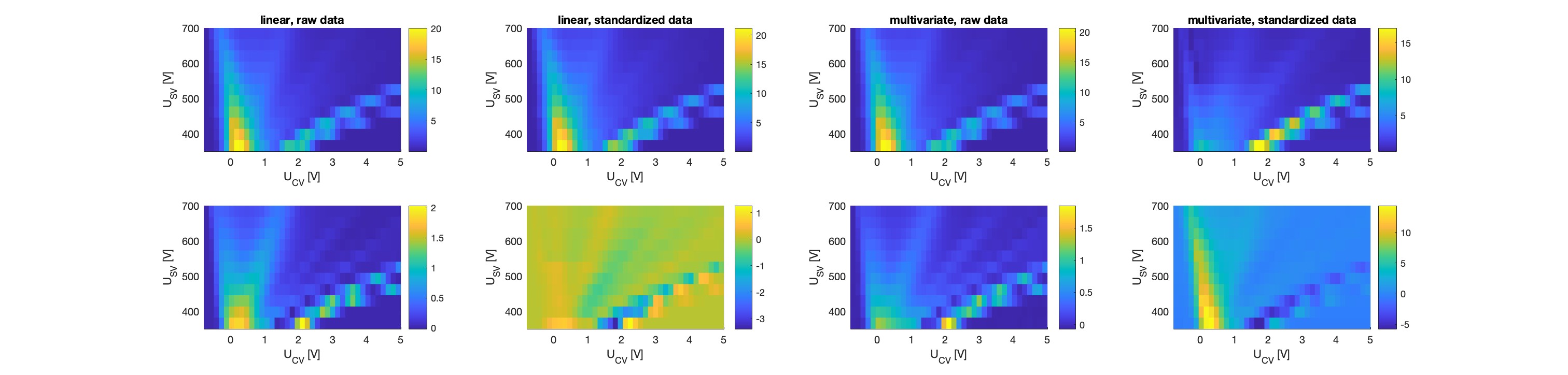}
\caption{
Pixelwise RMSEs for linear and multivariate regression models with raw as well as standardized DMS data from muscle tissue (upper row) and corresponding differences of pixelwise standard deviations and pixelwise RMSEs (lower row). Responses in pA.}
\label{fig:MusRes}
\vspace{-2ex}
\end{figure}

Figure~\ref{fig:AdiRes} (adipose tissue samples) and Figure~\ref{fig:MusRes} (muscle tissue samples) illustrate the superiority of employing multivariate regression models fitted to normalized data. In both figures RMSEs of estimated responses to CV/SV combinations are shown in the upper rows. In regions where the alpha curves are located, the observed RMSEs are considerably larger than in regions containing redundant responses. Such behavior was expected since considerably higher standard deviations were observed at these locations in the original data (compare Figures~\ref{fig:avg+sdAdi} and~\ref{fig:avg+sdMus}). One obvious difference is the magnitude of the RMSEs in regions containing the alpha curves. For the linear regression models with raw as well as normalized data, and the multivariate regression models with the raw data, the RMSEs are similar in these regions, reaching up to approximately 25\,pA. However, for the multivariate regression models with normalized data, the RMSEs are clearly lower, reaching up to approximately 16\,pA.


These differences between the four methods are further illustrated in the lower rows of Figures~\ref{fig:AdiRes} and~\ref{fig:MusRes}, which display differences between the standard deviation of a CV/SV combination and its corresponding RMSE. While the first three methods have RMSEs slightly below or above the corresponding standard deviations observed in the original data, the multivariate regression models with normalized data yield RMSEs well below the corresponding standard deviations. For the upper right corner, the models do not provide any improvement, which could be expected as at these CV/SV combinations ions are unable to pass the DMS filter and can be considered as noise~\cite{Anttalainen2019}. Similarly, no improvements for CV/SV combinations under the reactant ion peak could be expected due to the low separation field in combination with high CVs~\cite{Anttalainen2019}. For all alpha curve regions, except the region containing reactant ion peaks, the combination of multivariate regression and data normalization yielded RMSEs that are up to 14.4\,pA lower than the standard deviations. One shortcoming of the method is that for the reactant ion peaks of muscle tissue, the observed RMSEs are slightly higher (up to 5\,pA) than the corresponding standard deviations. This might be explained by the higher water content in muscle tissue as compared to adipose tissue. The vaporization of the tissue samples with laser also evaporates water in the sample. As the quantity of water vapor in muscle tissue is high, large fluctuations in the location and magnitude of the reactant ion peak could be expected (compare Figures~\ref{fig:avg+sdAdi} and~\ref{fig:avg+sdMus}). In future research, humidified air could be measured at varying temperatures and humidity levels to develop regression models that better capture the impact of water vapor on the location and magnitude of the reactant ion peak.


\section{Conclusion and outlook}

This paper showed, on the example of surgical smoke samples evaporated from porcine adipose and skeletal muscle tissue, that changes in absolute humidity and temperature considerably change differential mobility spectrometry measurements. The standard deviations for responses to specific combinations of compensation and separation voltages reached approximately one third of the average responses, indicating that locations of the so-called alpha curves in the dispersion plots changed. Alpha curves are the regions of interest when analyzing DMS data and could be used for identifying the tissue type based on its surgical smoke measured by DMS.

To better understand the dependencies on humidity and temperature, linear and multivariate regression models were fitted to DMS measurements collected at known humidity and temperature. While the linear regression models were fitted to DMS responses for fixed compensation and separation voltages, the multivariate regression models were fitted to DMS responses for fixed separation voltage and 40 different compensation voltages. This means, linear models were fitted to pixels and multivariate models were fitted to rows of the dispersion plots. In addition, the impact of row-wise normalized DMS measurements on the performance of both regression models were tested. The model coefficients were trained with part of the DMS data from adipose and muscle tissue, before being used for predicting DMS data for known tissue type, absolute humidity, and temperature. The prediction performance of both models trained with raw as well as row-wise normalized data was finally evaluated by comparing the predicted DMS data with the corresponding DMS measurements. Root mean square errors were only slightly smaller than the standard deviations when using the linear regression models trained with raw as well as normalized data or multivariate regression models trained with the raw data. Hence, these models in combination with the used data cannot be recommended for estimating DMS data of a known tissue type at known absolute humidity and temperature. 

However, for multivariate regression models trained on row-wise normalized data, the root mean square errors of predicted DMS measurements were up to 14.4\,pA smaller than the corresponding standard deviations in regions containing the alpha curves. Only for the reactant ion peak in dispersion plots of muscle tissue, the multivariate regression model trained on normalized data yielded root mean square errors exceeding the corresponding standard deviations by up to 5\,pA. Further tests on the impact of water vapor on the location and magnitude of the reactant ion peak will be carried out in the future. Nevertheless, this study confirmed that it is possible to predict the DMS responses for known tissue types at known absolute humidity and temperature, which means that it is possible to standardize DMS measurements rather than the conditions in which the measurements are collected.

One field of application for the tested methods is the analysis of surgical smoke during surgeries for distinguishing between, for example, healthy and cancerous tissue. As demonstrated in the literature, it is difficult to impossible to sustain constant humidity and temperature during surgeries. Even within a single surgery, for example, the water content in the surgical smoke can fluctuate. This presents a challenge in developing classification algorithms that accurately distinguish healthy from cancerous tissue. By standardizing DMS measurements used for training and validating the algorithms as well as standardizing DMS measurements during surgery, the accuracy of the classifiers could be improved tremendously. In the future, we will test the proposed regression method in combination with classifiers that proved to enable accurate tissue identification in controlled environmental conditions in an \textit{in vivo} study.


\section*{Acknowledgment}

This study was financial supported by the Research Council of Finland (P.M., A.Ra.: decision no. 360768; N.O.: decision no. 292477), Sigrid Juselius foundation (A.K., N.O., A.V.: decision no. 240148), the Doctoral School of Tampere University (A.K., M.L.), The Finnish Foundation for Technology Promotion (A.K.: grant no. 7671), Emil Aaltonen Foundation (A.K.: grant no. 210073K), The Finnish Medical Foundation (M.L.: grant no. 2167, 4038), Cancer Foundation of Finland (M.L.), Competitive State Research Financing of the Expert Responsibility Area of Tampere University Hospital and Pirkanmaa Hospital District (N.O.: grant no. 9AA057, 9x040, 9v044, 9T044, 9U042, 9s045, 150618, 151B03), Tampere Tuberculosis Foundation (N.O. ), and by the European Commission (Horizon 2020, A.V.: grant agreement 777222).

We thank Markus Karjalainen and Artturi Vuorinen for assisting in the measurements and sampling system hardware as well as Anna Anttalainen and Meri Mäkelä for extracting and formatting the data from the ENVI-AMC DMS unit. We also thank the personnel of Paija slaughterhouse (Paijan tilateurastamo, Urjala, Finland) for providing the sample material.

The study material used in the research was slaughterhouse offal. Ethical approval from the Ethics Committee of the Tampere Region was not needed for conducting this research.

\section*{Author contributions}

\begin{itemize}
\item Conceptualization: Philipp Müller, Anton Rauhameri
\item Methodology: Philipp Müller, Anton Rauhameri, Anton Kontunen, Antti Roine, Niku Oksala, Antti Vehkaoja, Maiju Lepomäki
\item Software: Philipp Müller, Anton Kontunen
\item Validation: Philipp Müller
\item Formal analysis: Philipp Müller
\item Investigation: Maiju Lepomäki
\item Resources: Anton Kontunen, Antti Roine, Niku Oksala, Antti Vehkaoja
\item Data curation: Anton Kontunen, Maiju Lepomäki
\item Writing – original draft preparation: Philipp Müller, Gary A. Eiceman
\item Writing – review and editing: Philipp Müller, Gary A. Eiceman; Anton Rauhameri, Anton Kontunen, Antti Roine, Niku Oksala, Antti Vehkaoja, Maiju Lepomäki
\item Visualization: Philipp Müller
\item Supervision: Antti Vehkaoja, Maiju Lepomäki
\item Project administration: Philipp Müller
\item Funding acquisition: Philipp Müller, Anton Kontunen, Antti Roine, Niku Oksala, Antti Vehkaoja, Maiju Lepomäki
\end{itemize}

\vspace{12pt}

\begin{thebibliography}{1}

\bibitem{Zamora2012}
D.~Zamora and M.~Blanco, ``Improving the efficiency of ion mobility spectrometry analyses by using multivariate calibration", {\it Analytica Chimica Acta}, vol. 726, 50--56, 2012.

\bibitem{Muller2019}
P.~Müller, K.~Salminen, A.~Kontunen, et al., ``Online scent classification by ion-mobility spectrometry sequences," {\it Front. Appl. Math. Stat.}, vol.  5, no. 39, 2019.

\bibitem{Dodds2019}
J.~Dodds and E.~Baker, ``Ion mobility spectrometry: fundamental concepts, instrumentation, application, and the road ahead," {\it J. Am. Soc. Mass Spec.}, vol. 30, no. 11, 2185--2195, 2019.

\bibitem{Anttalainen2019}
O.~Anttalainen, J.~Puton, A.~Kontunen, et al., ``Possible strategy to use differential mobility spectrometry in real time applications," {\it Int. J. Ion Mobil. Spec.}, vol. 23, 1--8, 2019.

\bibitem{Rauhameri2024}
A.~Rauhameri, A.~Robiños, O.~Anttalainen, et al., ``Classification of volatile organic compounds by differential mobility spectrometry based on continuity of alpha curves,'' {\it IEEE Access}, vol. 12, 130571--130582, 2024.

\bibitem{Kothawade2025}
G.S.~Kothawade, L.R.~Khot, A.K.~Chandel, et al., ``Feasibility of Little Cherry/X-Disease Detection in Prunus avium using Field Asymmetric Ion Mobility Spectrometry,'' {\it Sensors}, vol. 25, no. 7, 2034, 2025.

\bibitem{Kontunen2018}
A.~Kontunen, M.~Karjalainen, J.~Lekkala, et al., ``Tissue identification in a porcine model by differential ion mobility spectrometry analysis of surgical smoke," {\em Ann. Biomed. Eng.}, vol. 46, 1091--1100, 2018.

\bibitem{Karjalainen2018}
M.~Karjalainen, A.~Kontunen, S.~Saari, et al., ``The characterization of surgical smoke from various tissues and its implications for occupational safety," {\em PLoS One}, vol. 13, no. 4, e0195274, 2018.

\bibitem{Lepomaki2022}
M.~Lepomäki, A.~Anttalainen, A.~Vuorinen et al., ``Laser desorption tissue imaging with differential mobility spectrometry," {\it Exp. \& Molec. Pathology}, vol. 125, 104759, 2022.

\bibitem{Haapala2022}
I.~Haapala, A.~Rauhameri, A. Roine, et al., ``Method for the intraoperative detection of IDH mutation in gliomas with differential mobility spectrometry," {\it Current Oncology}, vol. 29, no. 5, 3252--3258, 2022.

\bibitem{Hermelo2025}
I.~Hermelo, I.~Haapala, M.~Mäkelä, et al., ``Patient-derived glioma organoids real time identification of IDH mutation, 1p/19q-codeletion and CDKN2A/B homozygous deletion with differential ion mobility spectrometry," {\it J. Neuro-Oncology}, vol. 171, 691--703, 2025.

\bibitem{Eiceman2014}
G.~Eiceman, Z.~Karpas, and H.~Hill jr., \emph{Ion mobility spectrometry}, 3rd~ ed., CRC Press, 2014.

\bibitem{Kontunen2021}
A.~Kontunen, U.~Karhunen-Enckell, M.~Karjalainen, et al., ``Tissue identification from surgical smoke by differential mobility spectrometry: an in vivo study," {\it IEEE Access}, vol. 9, 168355--168367, 2021.

\bibitem{Cheng2021}
M.-H.~Cheng, C.-H.~Chiu, C.-T.~Chen, et al., ``Sources and components of volatile organic compounds in breast surgery operating rooms," {\it Ecotoxicology and Environmental Safety}, vol. 209, 111855, 2021.

\bibitem{Chen2022}
C.-T.~Chen, S.-F.~Huang, C.-J.~Li, et al., ``Distribution of ultrafine aerosols and volatile organic compounds from surgical smoke during electrocauterization," {\it Air Quality, Atmosphere \& Health}, vol. 15, pp. 2009-2020, 2021.

\bibitem{Ieritano2021}
C.~Ieritano, J.~Campbell, and W.~Hopkins, ``Predicting differential ion mobility behaviour in silico using machine learning," {\it Analyst}, 146, 4737, 2021.

\bibitem{Iertiano2022}
C.~Iertiano and W.~Hopkins, ``The hitchhiker’s guide to dynamic ion–solvent clustering: applications in differential ion mobility spectrometry," {\it Phys. Chem. Chem. Phys.}, vol. 24, 20594, 2022.

\bibitem{Anttalainen2021}
A.~Anttalainen, M.~Mäkelä, P.~Kumpulainen, et al., ``Predicting lecithin concentration from differential mobility spectrometry measurements with linear regression models and neural networks," {\it Talanta}, vol. 225, 121926, 2021.

\bibitem{Weber1997}
L.M.~Weber and M.~Splei{\ss}, ``Formation of volatile organic compounds from peptides during CO$_2$-IR-laser irradiation of different mammalian tissues," {\it J. Anal. Appl. Pyrolysis}, vol. 39, 65--77, 1997.

\bibitem{Albrecht1994}
H.~Albrecht, R.~Hagemann, W.~Wäsche et al., ``Volatile organic components in laser and electrosurgery plume," {\it Proc. SPIE 2077}, 1994.

\bibitem{Spleiss1995}
M.~Splei{\ss}, Lothar.W.~Weber, T.H.~Meier, and B.~Treffler, ``Identification and quantification of selected chemicals in laser pyrolysis products of mamalian tisues," {\it Proc. SPIE 2323}, 1995.

\bibitem{Barrett2004}
W.L.~Barrett and S.M.~Garber, ``Surgical smoke—a review of the literature", {\it Bus. Brief: Glob. Surg.}, 1--7, 2004.

\bibitem{AlSahaf2007}
O.S.~Al Sahaf, I. Vega-Carrascal, F.O.~Cunningham et al., ``Chemical composition of smoke produced by high-frequency electrosurgery," {\it Ir. J. Med. Sci.}, vol. 176, 229--232, 2007.

\bibitem{Yeganeh2020}
A.~ Yeganeh, M.~Hajializade, A.P.~Sabagh et al., ``Analysis of electrocautery smoke released from the tissues frequently cut in orthopedic surgeries," {\it World J. Orthop.}, vol. 11, no. 3, 177--183, 2020.

\bibitem{Borsdorf2015}
H.~Borsdorf, P.~Fiedler, and T.~Mayer, ``The effect of humidity on gas sensing with ion mobility spectrometry," {\it Sensors Actuators, B Chem.} vol. 218, 184--190, 2015.

\bibitem{Safaei2019SciRep}
Z.~Safaei, G.~Eiceman, J.~Puton et al., ``Differential mobility spectrometry of ketones in air at extreme levels of moisture," {\it Sci Rep}, vol. 9, 5593, 2019.

\bibitem{Safaei2019ACA}
Z.~Safaei, T.~Willy, G.~Eiceman et al., ``Quantitative response in ion mobility spectrometry with atmospheric pressure chemical ionization in positive polarity as a function of moisture and temperature," {\it Analytica Chimica Acta}, vol. 1092, 144--150, 2019.

\bibitem{Mäkinen2011}
M.~Mäkinen, M.~Sillanpää, A.K.~Viitanen et al., ``The effect of humidity on sensitivity of amine detection in ion mobility spectrometry," {\it Talanta}, vol. 84, no. 1, 116--121, 2011.

\bibitem{Szczurek2017}.
A. Szczurek, M. Maziejuk, M. Maciejewska et al., ``BTX compounds recognition in humid air using differential ion mobility spectrometry combined with a classifier," {\it Sensors Actuators, B Chem.}, vol. 240, 1237--1244,  2017.

\bibitem{Krylova2003}
N.~Krylova, E.~Krylov, and G.~Eiceman, ``Effect of moisture on the field dependence of mobility for gas-phase ions of organophosphorus compounds at atmospheric pressure with field asymmetric ion mobility spectrometry", {\it J. Phys. Chem. A}, vol. 107, 3648--3654, 2003.

\bibitem{Kuklya2015}
A.~Kuklya, F.~Uteschil, K.~Kerpen, et al., ``Effect of the humidity on analysis of aromatic compounds with planar differential ion mobility spectrometry", {\it Int. J. Ion Mobil. Spec.}, vol. 18, 67--75, 2015.

\bibitem{Wolanska2023}
I.~Wola\'nska, K. Piwowarski, E.~Budzy\'nska, and J.~Puton, ``Effect of humidity on the mobilities of small ions in ion mobility spectrometry", {\it Anal. Chem.}, vol. 95, 8505--8511, 2023.

\bibitem{Krylov2009}
E.~Krylov, S.~Coy, and E.~Nazarov, ``Temperature effects in differential mobility spectrometry", {\it Int. J. Mass Spec.}, vol. 279, 119--125, 2009.

\bibitem{Virtanen2022}
J.~Virtanen, A.~Anttalainen, J.~Ormiskangas et al., ``Differentiation of aspirated nasal air from room air using analysis with a differential mobility spectrometry-based electronic nose: a proof-of-concept study," {\it J. Breath Res.}, vol. 16, 016004, 2022.

\bibitem{Huff2005}
E.~Huff-Lonergan and S. Lonergan, ``Mechanisms of water-holding capacity of meat: The role of
postmortem biochemical and structural changes," {\it Meat Science}, vol. 71, 194--204, 2005.

\bibitem{Wang1976}
J.~Wang and R.N.~Pierson, ``Disparate Hydration of Adipose and Lean Tissue Require a New Model for Body Water Distribution in Man," {\it The Journal of Nutrition}, vol. 106, no. 12, 1687--1693, 1976.

\bibitem{Thomas1962}
L.W.~Thomas, ``The chemical composition of adipose tissue of man and mice," {\it Q J Exp Physiol Cogn Med Sci.}, vol. 47, 179--188, 1962.

\end{thebibliography}
\end{document}